\documentclass[authoryear, 3p]{elsarticle}

\usepackage{epstopdf}
\usepackage{amsmath}
\usepackage{algorithm}
\usepackage[noend]{algpseudocode}
\usepackage{graphicx}
\usepackage{rotating}
\usepackage{mdframed}
\usepackage{url}

\algrenewcommand{\algorithmiccomment}[1]{#1}
\journal{Expert Systems with Applications}

\begin{document}

\begin{frontmatter}

\title{Detection of opinion spam based on anomalous rating deviation}

\author[rmit]{David Savage\corref{cauthor}}
\cortext[cauthor]{Corresponding author}
\ead{david.savage@rmit.edu.au}

\author[rmit]{Xiuzhen Zhang}
\ead{xiuzhen.zhang@rmit.edu.au}

\author[rmit]{Xinghuo Yu}
\ead{xinghuo.yu@rmit.edu.au}

\author[rmit,austrac]{Pauline Chou}
\ead{Pauline.Chou@austrac.gov.au}

\author[rmit]{Qingmai Wang}
\ead{qingmai.wang@rmit.edu.au}

\address[rmit]{School of CS\&IT, RMIT University, GPO Box 2476, Melbourne, Victoria, 3001, Australia}
\address[austrac]{Australian Transaction Reports and Analysis Centre, PO Box 13173, Law Courts, Melbourne, Victoria, 8010, Australia}

\begin{abstract}

The publication of fake reviews by parties with vested interests has become a severe problem for consumers who use online product reviews in their decision making. To counter this problem a number of methods for detecting these fake reviews, termed opinion spam, have been proposed. However, to date, many of these methods focus on analysis of review text, making them unsuitable for many review systems where accompanying text is optional, or not possible. Moreover, these approaches are often computationally expensive, requiring extensive resources to handle text analysis over the scale of data typically involved.

In this paper, we consider opinion spammers manipulation of average ratings for products, focusing on differences between spammer ratings and the majority opinion of honest reviewers. We propose a lightweight, effective method for detecting opinion spammers based on these differences. This method uses binomial regression to identify reviewers having an anomalous proportion of ratings that deviate from the majority opinion. Experiments on real-world and synthetic data show that our approach is able to successfully identify opinion spammers. Comparison with the current state-of-the-art approach, also based only on ratings, shows that our method is able to achieve similar detection accuracy while removing the need for assumptions regarding probabilities of spam and non-spam reviews and reducing the heavy computation required for learning.

\end{abstract}

\begin{keyword}
anomaly detection \sep binomial regression \sep classification  \sep online product reviews \sep opinion spam \sep review spam
\end{keyword}

\end{frontmatter}


\section{Introduction}

Online product reviews, reporting others experience with a given product, can be extremely useful for consumers making purchasing decisions. Given the bounty of choice available in online stores, product reviews provide a helpful aid for consumers attempting to gauge product quality and decide between different brands and different product models. However, in recent years, opinion spam, consisting of fake reviews published by individuals with vested interests, has become a major problem for consumers \citep{Jindal:2008,Kugler:2014,Heydari:2015}. Opinion spam typically involves the publication of fake product reviews for the explicit purpose of influencing a buyers' perceptions of quality and utility \citep{Jindal:2008,Kugler:2014}. By publishing numerous fake reviews, opinion spammers attempt to artificially inflate consumers' confidence that previous buyers are satisfied with their purchase. Alternatively, spammers may attempt to create an artificial belief that previous buyers of a competitors' product have come to be dissatisfied with their purchase.

Previous approaches to detecting opinion spam have tended to focus on analysis of review text (see \citet{Heydari:2015} for a comprehensive survey of existing methods). These approaches rely on the identification of duplicated passages of text occurring in multiple reviews  (e.g. \citet{Jindal:2008,Lau:2011,Mukherjee:2012,Mukherjee:2013}), or consider multiple text-based features, using manually identified opinion spam to train classifiers (e.g. \citet{Ott:2011,Li:2011,Ramkumar:2010,Fusilier:2014}). While these text-based approaches have been used with success, they suffer three major drawbacks \citep{Akoglu:2013}. First, detection of repeated text requires expensive comparisons, and without first narrowing down the selection of candidates the number of comparisons required may quickly become infeasible. Second, new training data is often required for different product domains (e.g. hardware products vs restaurant reviews), and third, manual identification of opinion spam for use in training can be an expensive and time-consuming undertaking. Moreover, many rating systems in use today require only a rating (typically expressed as a binary good/bad or as $1 - 5$ stars), with a text-based review optional (e.g. the Apple App Store), or not possible at all (e.g. the Facebook `like' system). Thus, there is a need to develop methods for detecting opinion spam based solely on ratings \citep{Akoglu:2013}. 

Many online shopping services display the mean rating for available products, and this has been shown to be a key piece of information used by consumers in making their purchasing decisions \citep{Chevalier:2006}. Thus, one way in which opinion spammers attempt to alter consumers' perception of quality is to manipulate the mean rating for a target product. By generating multiple reviews that appear to have originated from different users, spammers are able to significantly distort the mean rating \citep{Akoglu:2013,Mukherjee:2012}. However, in doing so, spammers are often required to post ratings that are at odds with those of honest reviewers, and consequently opinion spammers can be expected to have an abnormal number of reviews that significantly differ from the mean rating.

In this paper we propose a novel method for detecting opinion spammers attempting to manipulate mean ratings. This method significantly differs from previous approaches in two main ways. First, our method characterises reviewer behaviour in a manner that allows detection of opinion spammers using only ratings, without resorting to text-based analysis. Second, we fit a binomial model to the target set of reviews, and identify spammers as those reviewers exhibiting anomalous behaviour under this model. Consequently, our method accurately reflects the observed patterns of reviewer behaviour for the particular system under consideration, rather than relying on assumed parameter values describing this behaviour.

To demonstrate the utility of our approach we conduct experiments using both real and synthetic data, showing that our method can successfully identify opinion spammers. We compare our approach with the FraudEagle algorithm presented in \citep{Akoglu:2013}, which is also capable of detecting opinion spammers based only on ratings. FraudEagle has previously been shown to outperform alternative approaches and can therefore be considered the current state of the art for detection of the type of spammer discussed in this paper. Using a combination of real and synthetic data we demonstrate that our approach is able to achieve a similar level of performance to FraudEagle while providing a conceptually deeper, but computationally simpler, characterisation of spammer behaviour.

\section{Related Work}

Previous approaches for detection of opinion spam have typically involved supervised or unsupervised learning based largely on text-based features \citep{Heydari:2015}. While many of these approaches include some non-text-based features, the major focus to date has been on features such as n-gram counts and cosine similarity. Using these types of features, a wide variety of supervised and semi-supervised classifiers have been described \citep{Li:2011,Ott:2011,Lim:2010,Fusilier:2014}. These classifiers are able to successfully identify spam with a high degree of accuracy, however, in order to perform required training, these studies rely on manual labelling of reviews by domain experts, which is a time-consuming and costly endeavour.

To overcome the difficulties associated with manual labelling, an unsupervised approach has been proposed that applies an unsupervised Bayesian framework to detection of opinion spammers \citep{Mukherjee:2012,Mukherjee:2013}. In this framework, the \textit{spamicity} of each reviewer is modelled as a latent variable in a hierarchical model including both text- and non-text-based features. Experiments with real data sets showed that this approach is able to accurately identify opinion spammers, with posterior analysis suggesting that discrimination between spammers and non-spammers is largely driven by text-based features. However, we note that this posterior analysis also showed that rating deviation was an important aspect of spammer behaviour, and suggested that rating deviation may be more useful in separating spammers from non-spammers than consideration of early reviews and reviews consisting of extreme ratings.

While the vast majority of previous work has focused on text-based features, one exception is the FraudEagle algorithm \citep{Akoglu:2013}, which has been shown to successfully detect of opinion spammers using only product ratings. The FraudEagle algorithm uses a graph-based representation of the product-review system, with reviews represented as edges between reviewers and products. FraudEagle applies an iterative approach to spammer detection, whereby the inter-dependency between perceived product quality and the \textit{spamicity} of reviewers is resolved by updating scores for a given vertex, and then propagating this update along edges in the graph, converging when the scores for each vertex becomes consistent with its neighbours' scores. A similar approach is also proposed in \citet{Wang:2012}, however this study deals with reviews of retail stores, which can change in quality over time, requiring the timing of reviews to be taken into account by the detection algorithm.

In detecting opinion spammers, the FraudEagle algorithm relies on a set of parameters describing the different behaviour of honest reviewers and spammers. These parameters are difficult to estimate \textit{a priori}, and are consequently set to arbitrary values \citep{Akoglu:2013}. In this paper we take a significantly different approach to that of FraudEagle, which eliminates the requirement for these parameters, and does not require a graph representation of the system. Instead a binomial model of reviewer behaviour is fit to the target set of product ratings, resulting in a more accurate representation of reviewer behaviour, and at the same time greatly reducing the computational requirements.\\

In addition to those works focusing on opinion spam, related work is also found across a wide range of problem domains through the shared use of statistical anomaly detection. For example, statistical anomaly detection has previously been employed for detecting unusual movement in crowds \citep{Kratz:2009}, network intrusion \citep{Eskin:2000}, spam phone calls through VoIP \citep{Jung:2012}, and threats to operating system security \citep{Kruegel:2003}. A comprehensive review of anomaly detection in general is given in \citet{Chandola:2009}, with statistical anomaly detection discussed alongside alternative methods. Similar approaches are also discussed by \citet{Markou:2003} in their review of statistical novelty detection.

Our approach to detection of opinion spam focuses on deviation from the majority opinion. As we will discuss in section \ref{AlgorithmDescription}, we use the mean rating as a measure of majority opinion, as this is the measure most often shown to consumers. However, alternative measures of majority opinion have previously been studied in the context of decision making and decision theory (see for example \citet{Pasi:2006} and \citet{Vanivcek:2009}).

\section{Detection of opinion spammers through consideration of majority opinion}
\label{AlgorithmDescription}

The proposed method for detecting opinion spammers stems from two overarching assumptions regarding reviewer behaviour. (1) The majority of reviews are posted by honest reviewers, and as a consequence, distributions taken over large samples of the reviewer population will overwhelmingly reflect the behaviour of honest reviewers. (2) Honest reviewers will typically have similar expectations and perceptions of quality, such that the set of reviews for a given product will tend to exhibit a small degree of variance.

\begin{figure}
\centering
\includegraphics[width=0.49\textwidth, trim=0.15cm 0cm 1.2cm 2cm, clip]{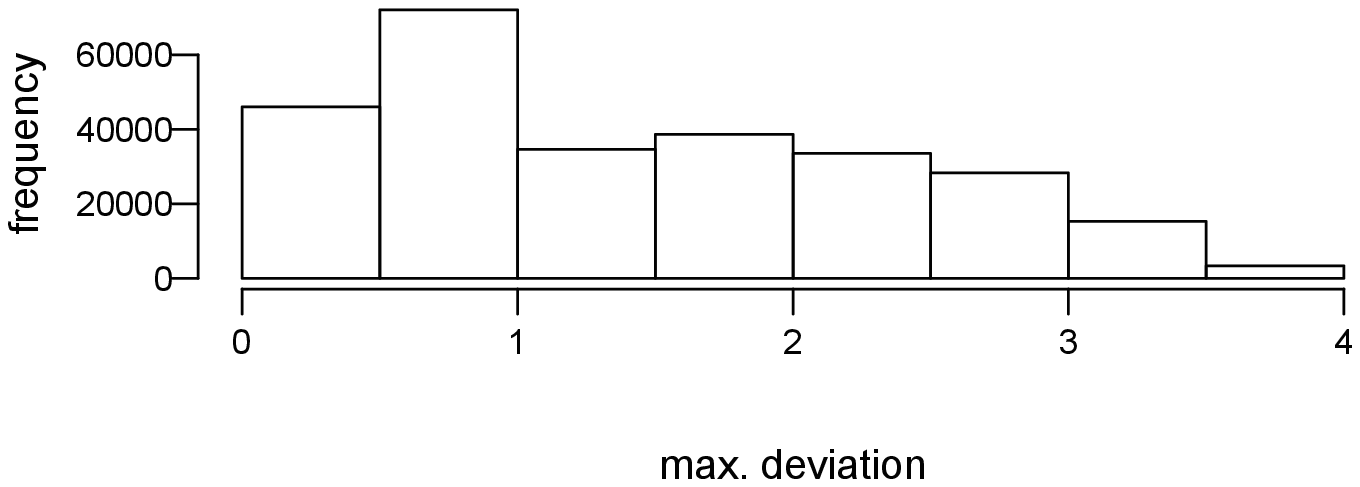}
\includegraphics[width=0.49\textwidth, trim=0.5cm 0cm 1.2cm 2cm, clip]{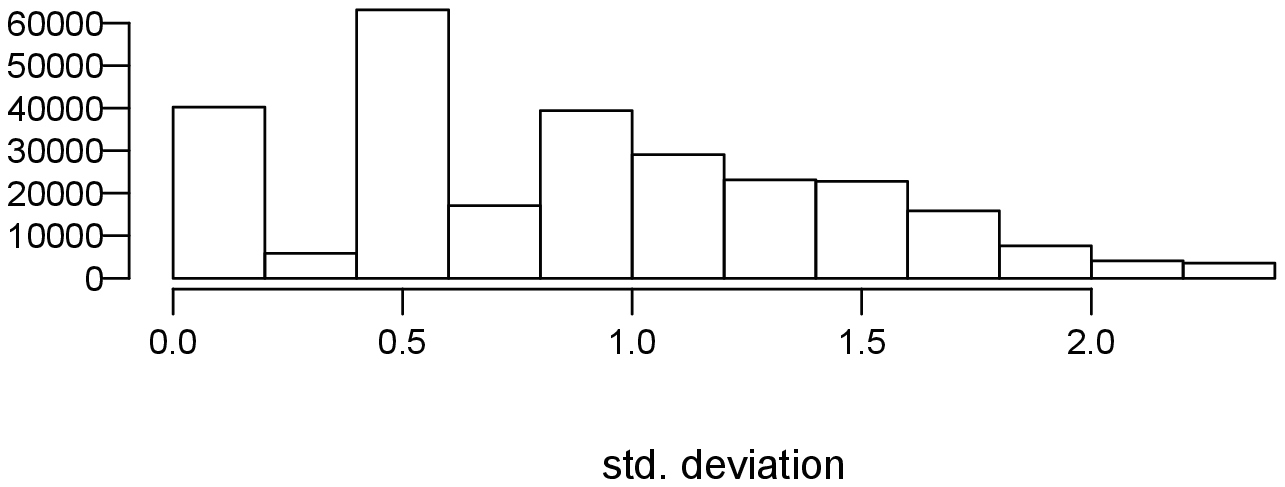}
\caption{Distribution of maximum and standard deviation from the mean rating for the Amazon data set.}
\label{VarDist}
\end{figure}

We take assumption 1 as given, since if this is not the case then the whole system of online peer reviews is completely broken. As justification for assumption 2,  Figure \ref{VarDist} shows the distribution of the standard deviation, and also the maximum deviation from the mean rating over approximately 272,000 products reviewed in the Amazon online store. For the majority of products, review scores appear to be relatively tightly clustered around the mean, rather than spread across the possible range of 5 stars.

Based on Figure \ref{VarDist}, we suggest honest reviewers tend to come to a similar opinion about a product, so that reviews that disagree with the mean rating will be relatively infrequent. Clearly such reviews may exist for legitimate reasons, for example the purchaser may happen to receive a defective version of an otherwise quality product, or a particular book or film generates a highly polarised response. However, it is unlikely that the same purchaser will consistently receive faulty models or review polarising products, therefore, we suggest that having a high proportion of reviews that disagree with the mean rating can be considered as a strong indicator that a reviewer may be a spammer.

Consider a product $p$ targeted by an opinion spammer (or a group of opinion spammers), who wishes to influence the average rating of the product. In manipulating the average rating, the spammer attempts to create an overall perception of either satisfaction or dissatisfaction on the part of previous purchasers. Assuming a 5-star rating system, we can model this manipulation by taking the midpoint rating $\lambda = 3$, and assuming that products having a mean rating $\overline{\sigma_p} \ge 3$ will be perceived as good while those having $\overline{\sigma_p} < 3$ will be perceived as bad. The spammers objective is to drive the mean rating in a particular direction so that the product in question will be perceived as good or bad by prospective purchasers. For a given reviewer $r$, a published rating $\sigma_{r,p}$ can be considered as agreeing ($\sigma_{r,p} \ge 3$ and $\overline{\sigma_p} \ge 3$ or $\sigma_{r,p} < 3$ and $\overline{\sigma_p} < 3$) or disagreeing ($\sigma_{r,p} \ge 3$ and $\overline{\sigma_p} < 3$ or $\sigma_{r,p} < 3$ and $\overline{\sigma_p} \ge 3$) with the majority opinion. Note that while we assume a 5-star system, our method could equally be applied to a binary system, taking $\lambda = 0.5$ as the midpoint and considering a mean rating $\overline{\sigma_p} \ge 0.5$ indicative of a good product and $\overline{\sigma_p} < 0.5$ as indicative of a bad product. Depending on the particular situation, we may also elect to set $\lambda$ to an alternative value. For example, we might consider $\lambda = 3.5$ to more accurately represent the point where consumers become far more likely to purchase particular products. Alternatively, we may assume that spammers are willing to post ratings of 3-stars in order to negate 5-star ratings of a product, attempting to drag the mean rating down without being too obvious. In this case, we may elect to use $\lambda = 4$ so that these types of spammers would still be detected.

In this paper we consider mean rating as our measure of majority opinion. This is because the mean rating is displayed to consumers in many online stores (e.g. Amazon, Apple App Store, Google Play, see also \citet{Chevalier:2006}). We therefore reason that spammers' manipulation of consumer perception requires them to alter the mean rating. For a particular situation, consideration of alternative measures, such as the median rating, may be beneficial. As described below, our approach considers a weighted mean, thus alternative measures that can be naturally weighted may be substituted for the mean in the approach described in this paper.

Using the rating model outlined above, we can determine the proportion of reviews for a given reviewer $r$ that disagree with the mean rating for their respective products. Whether or not we consider this proportion to be excessively high depends on how often we expect a random honest reviewer to post such a review. Since the overwhelming majority of reviewers are assumed to be honest, an estimate of this frequency can be easily calculated from the available data by simply considering the proportion of reviews that disagree with the mean rating across all available observations $\phi = n_{D} / n$. Taking $\phi$ as the probability that a random review $\sigma_{r,p}$ will disagree with the mean, we can estimate the \textit{spamicity} for each reviewer using a binomial distribution.

A binomial distribution models the outcome of independently repeating a random process a set number of times, where the random process results in a binary value $success$ or $ failure$. The distribution can be used to determine whether or not the observed proportions of $success$ and $failure$ across the repeated trials differ significantly from the expected proportions, given a known probability of obtaining $success$. In applying the binomial distribution to opinion spam, we treat the number of reviews posted by a given reviewer $n_r$ as the number of trials, and the number of reviews that disagree with the mean rating $k_r$ as the number of trials having an outcome $sucess$. We then calculate the probability $\psi_{r}$ of observing $k_{r}$ or more disagreeing reviews out of $n_{r}$ by random chance alone, taking $\phi$ as the probability of $success$.

\begin{eqnarray*}
\psi_{r} &= & P(X \ge k_r; n_r, \phi)\\
& = & 1 - P(X < k_r; n_r, \phi)\\
& = & \sum_{i = 0}^{k _{r}- 1}{\binom{n_{r}}{i}\phi^{i}(1 - \phi)^{n_{r} - i}}
\end{eqnarray*}

\noindent We can consider $\psi_{r}$ to be a measure of the reviewers honesty. A value of $\psi_r$ that is close to one indicates that the proportion of disagreeing reviews is within the expected bounds, while a value close to zero indicates that the proportion of disagreeing reviews is unexpectedly high. An estimate of the reviewers \textit{spamicity} can then be calculated as $s_{r} = 1  - \psi_{r}$.

In calculating the number of reviews that disagree with the majority it is important to remember that the mean rating reflects both the contributions by honest reviewers, and the deliberate manipulations by spammers. If a spammer successfully drives the mean rating above or below the midpoint, then honest reviewers would appear to disagree with the majority opinion, and may consequently be considered by the algorithm as candidate spammers \citep{Akoglu:2013,Wu:2010}. To correct for this situation, we can apply an iterative process whereby the contribution from each reviewer to the average rating for each product is successively reduced based on their proportion of non-majority reviews.\\

\noindent The iterative process used for correcting mean ratings begins by assuming that all reviewers are honest, and defining $u_{r,i}$ to be an estimated measure of whether reviewer $r$ is honest (i.e. not a spammer) derived from the $i^{th}$ iteration. For each reviewer we initialise $u_{r,0} = 1$. In each iteration $i > 1$ the mean rating $\overline{\sigma}_{p,i}$ for each product $p$ is calculated as a weighted arithmetic mean, with the contribution of each rating weighted using the values of $u_{r,i-1}$ from the previous iteration. Using this updated mean rating, the weights for the next iteration $u_{r,i}$ are calculated as $u_{r,i} = 1 - \frac{d_{r,i}}{n_{r}}$ where $d_{r,i}$ refers to the number of reviews for reviewer $r$ that disagree with the weighted mean rating for the respective products in iteration $i$ and $n_{r}$ is the total number of reviews published by that reviewer. The iterative process stops when a maximum number of loops have occurred or the observed change in the proportion of non-majority reviews becomes less than some threshold $\tau$ for all reviewers. Once the iterative process is complete, a binomial test is applied to each reviewer as described above.

For the experiments described in this paper, we set the threshold parameter $\tau = 1 \times e^{-5}$ and used a maximum number of 10 iterations. We found that 10 iterations was enough for the algorithm to effectively correct mean ratings, and often converged well before reaching this maximum. However, in some situations, updating the estimated honesty score for a small number of reviewers caused the weighted means of some products to flip back and forth from $> 3$ to $< 3$. Consequently, the proportion of reviews that disagree with the majority did not converge for these reviewers, and the maximum number of iterations was reached.

A step-by-step description of our approach is given as Algorithm 1. Lines $3  - 13$ relate to the iterative process used to correct the mean rating. Lines $5 - 7$ calculate the weighted mean for each product, and lines $9  - 12$ calculate the proportion of reviews that disagree with this updated mean. Lines $15  - 18$ relate to the assignment of a \textit{spamicity} score using the binomial test.

The presented algorithm scales linearly as a function of the number of reviewers and the number of reviews, $O(i_{max}(n_{reviews} + n_{reviewers}) + n_{reviewers})$, where $i_{max}$ is the number of iterations completed. Although we calculate the mean rating for each product, the total number of operations is dependent on the number of reviews for each of these products, thus this part of the algorithm scales with the number of reviews. We then calculate the proportion of non-majority reviews, so that the number of operations depends on the number of reviewers. Since we employ an iterative approach we multiply the number of operations by the number of iterations, and finally add the number of operations required for the binomial test, again a function of the number of reviewers.\\

\begin{algorithm}[t!]
\label{Algorithm}
\begin{algorithmic}[1]
\State{\textbf{Input:} reviewers, products, reviews, $\lambda$, $\alpha$, $\tau$, \textit{maxIterations}}
\State{\textbf{Output:} for each reviewer $r$, a \textit{spamicity} score $s_{r}$}
\Statex{}
\ForAll{$r$} \Comment{$u_{r,0} \leftarrow 1$ \hskip2em \textit{\#Assume all reviewers are honest}}
\EndFor
\Statex{}
\For{$i$ in $[1, maxIterations]$}
	\ForAll{products $p$} \Comment{\hskip2em \textit{\#Calculate the weighted mean for each product}}
		\State $R_{p} \leftarrow reviewers(p)$
		\State $\overline{\sigma}_{p} \leftarrow \sum_{r \in R_{p}}{\frac{\sigma_{r,p} u_{r,i}}{|R_{p}|}}$
	\EndFor
	\Statex{}
	\ForAll{reviewers $r$} \Comment{\hskip2em \textit{\#Calculate proportion of disagreeing reviews}}
		\State{$d_{r,i} \leftarrow count(\sigma_{r,p} \in reviews(r); \sigma_{r,p} < 3 \land \overline{\sigma}_{p,i} \ge 3 \vee \sigma_{r,p} \ge 3 \land \overline{\sigma}_{p,i} < 3 )$}
		\State $u_{r, i} \leftarrow 1 - \frac{d_{r,i}}{n_r}$ 
	\EndFor
	\Statex{}
	\If{$|u_{r,i-1}  - u_{r,i}| < \tau$ $\forall i$} \Comment{\textit{break} \hskip1em \textit{\#Break if convergence is achieved}}
	\EndIf
\EndFor
\Statex{}

\ForAll{reviewers $r$}
	\State $k_{r} \leftarrow count(\sigma_{r,p} \in reviews(r); \sigma_{r,p} < 3 \land \overline{\sigma}_{p} \ge 3 \vee \sigma_{r,p} \ge 3 \land \overline{\sigma}_{p} < 3 )$
	\State $\psi_{r} \leftarrow P(X \ge k_{r}; n_{r}, \phi)$ \Comment{\hskip5em \textit{\#Apply the binomial test}}
	\State $s_{r} \leftarrow 1 - p_{r}$
	
\EndFor
\end{algorithmic}
\caption{Algorithm for detecting opinion spammers from review ratings in a 5-star system. Inputs for the algorithm are the set of products, reviewers and reviews in the form of product ratings, a required significance level for the binomial test $\alpha$, the maximum number of iterations to be run, and the tolerance threshold $\tau$ for terminating the iterative process. Outputs are a $p$-value (probability reviewer is honest) for each reviewer and a label \textit{honest} or \textit{spammer} based on the calculated $p$-value and the given significance level}
\end{algorithm}

\section{Evaluation}

Evaluation of our method consisted of comparison with the recently proposed FraudEagle algorithm \citep{Akoglu:2013} and application of our approach to a real data set consisting of product reviews published on Amazon.com. The FraudEagle algorithm is a network-based approach to the detection of opinion spam that considers network structure as well as rating scores as part of an iterative message passing algorithm. FraudEagle has been evaluated against synthetic and real data sets, and has been shown to outperform a number of similar network-based approaches.

\subsection{Comparison with FraudEagle using synthetic data}

FraudEagle models product reviews as a bipartite graph, with reviewers and products represented as vertices and reviews as edges linking reviewers to products. Reviews of products are assumed to have a binary rating, \textit{good} or \textit{bad}. Products are similarly given a binary label, \textit{good} or \textit{bad}, depending on the number of \textit{good} and \textit{bad} reviews. FraudEagle also assigns a binary label to each reviewer, being either \textit{honest} or \textit{fraudulent}. FraudEagle takes a set of a parameters describing the probability that a reviewer with a given label will give a \textit{good} or \textit{bad} review of a product that is inherently \textit{good} or \textit{bad} (e.g. the probability that an \textit{honest} reviewer will give a \textit{good} product a \textit{bad} review). FraudEagle applies an iterative message passing algorithm, whereby the labels on products and reviewers are updated in a manner that maximises the likelihood of the network configuration (i.e. which reviewers and products have which labels) under the given probabilities. In this way, FraudEagle also considers deviation from majority opinion and attempts to discount the contribution of likely spammers. See \citet{ Akoglu:2013} for further details.

To compare our approach with FraudEagle we generated two synthetic data sets using two different models of spammer behaviour. Since FraudEagle considers product reviews as a bipartite graph, these data sets were generated using a random graph generator, RTG \citep{Akoglu:2009}. FraudEagle has been previously evaluated using synthetic data produced using RTG, thus we view RTG as a suitable generator for our comparison. Using RTG we generated two sets of random bipartite graphs, and for each graph we then generated a rating score for each edge $(v_i, v_j)$, representing a review of product $j$ by reviewer $i$. In generating these ratings we applied two different models of spammer behaviour, described below, giving two test sets of synthetic data. For both of these test sets, we applied our approach and the FraudEagle algorithm to all graphs, combining the results within each set. We then plotted a receiver operating characteristic (ROC) curve for both sets, using the area under the curve (AUC) as a metric for performance.

In the context of this paper, the false positive rate for the ROC curve refers to honest reviewers that are misidentified as spammers, while the true positive rate refers correctly identified spammers. Both our approach and FraudEagle calculate a \textit{spamicity} score for each reviewer, and reviewers having a \textit{spamicity} score above some threshold are assumed to be spammers. By varying this threshold, different false positive rates and true positive rates may be achieved. The ROC curve shows the effect of varying this threshold on the accuracy of each approach. For a given rate of false positives (misidentified honest reviewers), the ROC curve shows the expected rate of true positives (correctly identified spammers). In attempting to counter opinion spam, the cost of false positives is likely to be higher than false negatives, as flagging accounts honest reviewers as spam is likely to cause offence and significant public backlash. Consequently, performance at a low rate of false positives is important for detection of opinion spam.\\

Each of the synthetic test sets consisted used for evaluation consisted of 30 random bipartite graphs generated using RTG. We generated multiple graphs for each data set to minimise the possibility of skewed results stemming from the use of a random process. Previous evaluation of the FraudEagle algorithm also used RTG generator, and we used the same parameter values as this evaluation, $W = 5000$, $k = 5$, $q = 0.4$ and $\beta = 0.6$ (see \citet{Akoglu:2013} and \citet{Akoglu:2009} for details). The mean edge count for the resulting graphs was 1359. For each of the 30 graphs, we deemed the 7 products with the highest degree to be `famous'. The purpose of the famous products was to simulate spammers attempts to disguise their behaviour, by publishing honest reviews of well known products.

For the first test set, we followed the same process outlined in \citet{Akoglu:2013} for modelling spammer behaviour. In each graph we assigned the inherent value of all products to be `good' and then randomly selected 4 reviewers to represent spammers. Ratings posted by the selected spammers were set to `bad' unless the review happened to be directed towards one of the famous products, in which case it was set to `good'. For the remaining reviewers, deemed to be honest, all ratings were set to `good'.

For our second test set we applied an alternative model of spammer behaviour, which we believe is a more realistic representation of spammer behaviour. For this test set we calculated the distribution of ratings from the available set of Amazon product reviews, and using this distribution sampled a random rating for each edge in each graph, representing the generated reviews. After rating scores had been sampled for all edges, we randomly selected five reviewers to represent spammers, and the ratings for these five reviewers were flipped, so that, for example, an initial rating of 5 would be transformed to a 1, a rating of 2 would be transformed to 4, while a rating of 3 would remain unchanged. We reason that flipping the scores in this way simulates spammers posting reviews that attempt to drive the average rating away from the majority opinion. Because FraudEagle only considers a binary rating system we treated ratings of $\sigma_{r,p} \ge 3$ to be `good' and ratings of $\sigma_{r,p} < 3$ to be `bad' as described in \citet{Akoglu:2013}.

For each of the test sets we ran our iterative process on each network and then performed the binomial test, obtaining a \textit{spamicity} score for each reviewer. FraudEagle was also run for each network in each test set. Within each test set we combined results from all 30 graphs, giving two sets of \textit{spamicity} scores, one for each test set. For each of these two sets we then calculated an ROC curve based on the \textit{spamicity} scores reported by our approach and by FraudEagle.

\begin{figure}
\centering
\includegraphics[width=0.49\textwidth]{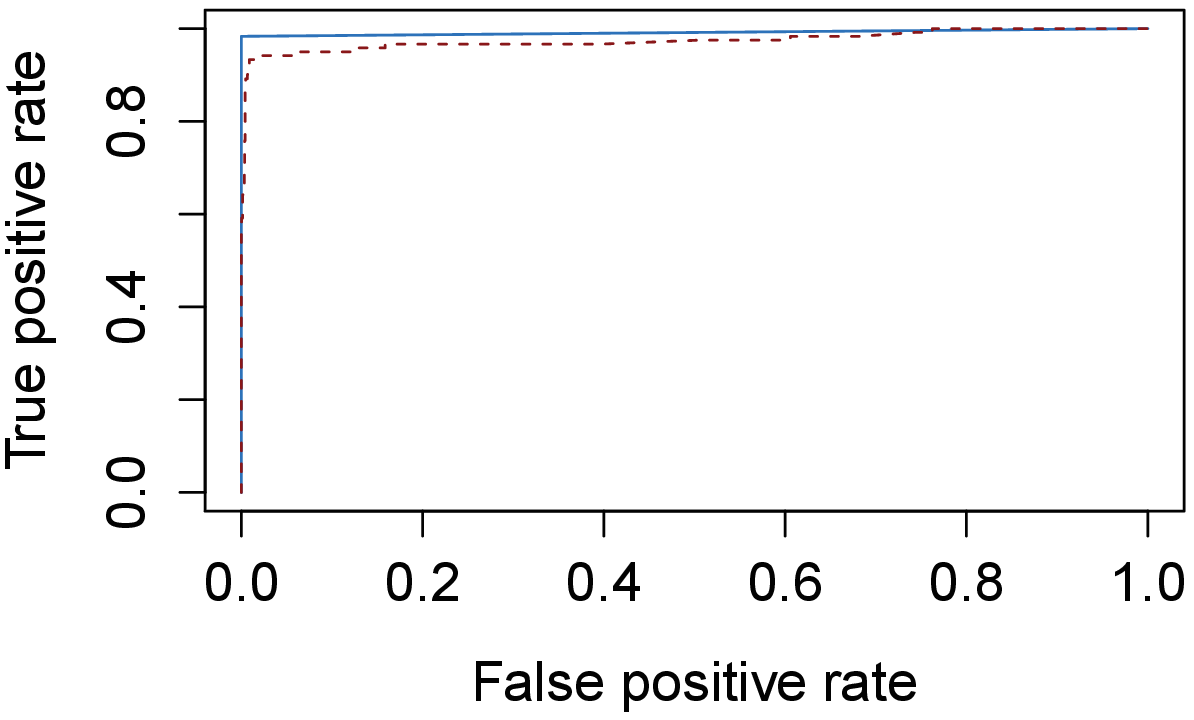}
\includegraphics[width=0.49\textwidth]{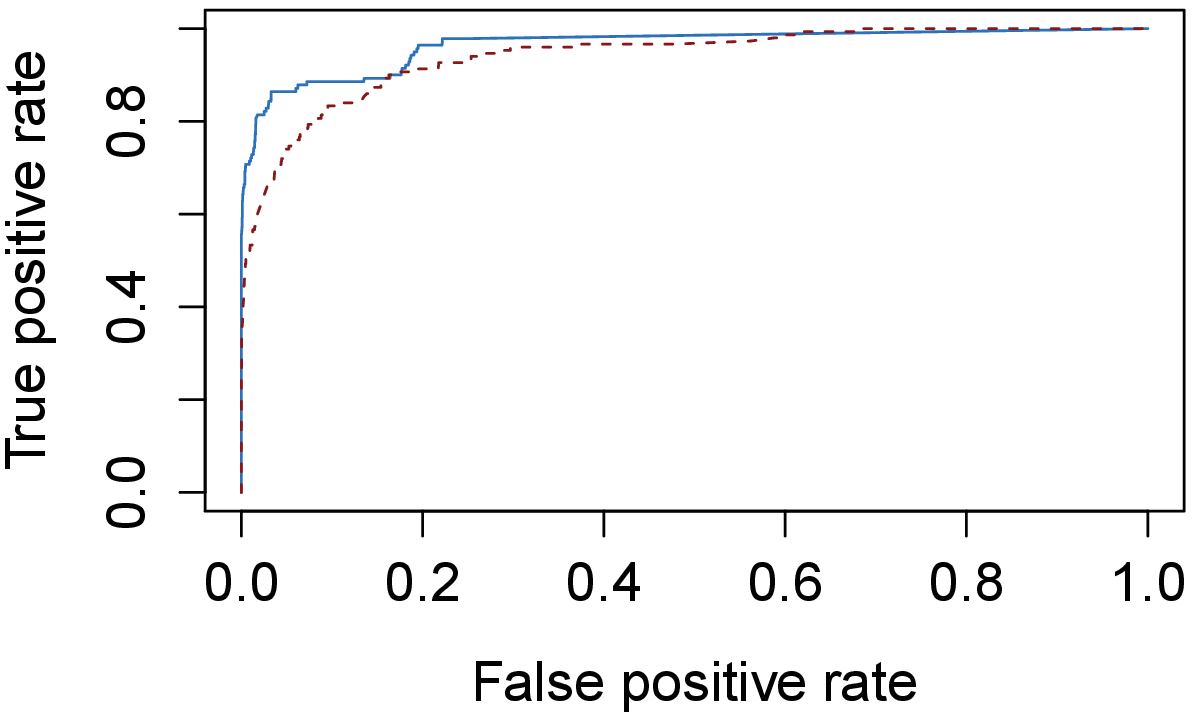}
\caption{ROC for our approach (blue, solid) and FraudEagle (red, dashed) applied to synthetic data sets using two different models of spammer behaviour. The left panel represents a binary review system where all products are good, honest reviewers always give good ratings and spammers always give bad ratings unless reviewing a `famous' product. The AUC is 0.964 for our approach and 0.940 for FraudEagle. The panel on the right represents behaviour where honest review ratings were drawn from the global distribution and spammer review ratings were then flipped so that any ratings greater than 3 were transformed to be less than 3 and vice versa. The AUC is 0.992 for our approach and 0.975 for FraudEagle.}
\label{ROC}
\end{figure}

Results of our comparison with the FraudEagle algorithm are given in Figure \ref{ROC}, showing the respective ROC curves. These results show that our approach out-performed FraudEagle on both sets of synthetic data, with our approach achieving a higher AUC. While this improvement over FraudEagle is marginal for both data sets, considering the simplicity of our approach compared to the FraudEagle algorithm, we believe that this comparison clearly demonstrates the viability of our approach. We also note that the ROC curve shows that our approach achieves a high rate of true positives for low rates of false positives; an important characteristic given the high cost of false positives in detection of opinion spam.

\subsection{Application to Amazon product reviews}
\label{DataSet}

We downloaded the Amazon product review data set from \url{http://liu.cs.uic.edu/download/data/} (password required, a link is provided on this page to obtain access, please see \citet{Jindal:2008}). After removing all reviews with missing fields, this data set consists of 5,838,041 reviews of 1,230,915 products, published by a total of 2,146,057 reviewers.

To ensure that our data contained a reasonable set of connected reviewers and products, we constructed a bipartite network from the product review data and extracted the largest connected component. We then removed all reviewers from this component having less than 3 or more than 5000 reviews, and all products having only a single review or more than 1000 reviews. We reason that reviewers having more than 5000 reviews are already somewhat suspicious and those having less than 3 reviews require different methods to detect (e.g. clusters of single positive review, see \cite{Wu:2010}). For products having more than 1000 reviews, the mean rating is less likely to be strongly influenced by fake reviews. Products having a single review obviously prevent any deviation from the majority. After removing products, some reviewers can become orphaned, therefore we removed reviewers having zero reviews as a final pre-processing step. The final data set consisted of 5,018,344 reviews of 570,606 products by 1,859,242 reviewers.\\

To evaluate the performance of our approach, we ran the iterative process described in section \ref{AlgorithmDescription} on the Amazon data set. We then applied the binomial test using a significance value of $\alpha = 0.05$, and selected those reviewers having a significant number of non-majority reviews as candidate spammers. Since we apply the test multiple times (once for each reviewer), we used a Bonferonni correction for the significance level, $\alpha = 0.05 / n_{reviewers}$. After applying the binomial test, we then further filtered the candidate spammers by removing those candidates having more than 50 reviews. This process resulted in 187 reviewers being identified as candidate spammers. 

Evaluation of candidate spammers was performed by considering a set of alternative features shown in other studies to be strong indicators of opinion spam \citep{Jindal:2008,Mukherjee:2012,Mukherjee:2013,Akoglu:2013}. In particular, we considered features relating to content similarity (repeated review text),  numerous extreme ratings (1-star or 5-star), and posting of multiple reviews on the same day. Note that calculation of content similarity is possible in our example data-set, but in many rating systems, review text would not be present and this feature could not be relied on for identification of spammers.\\

\begin{figure}
\centering
\includegraphics[width=0.3\textwidth, trim=0cm 0.8cm 1cm 1cm, clip=true]{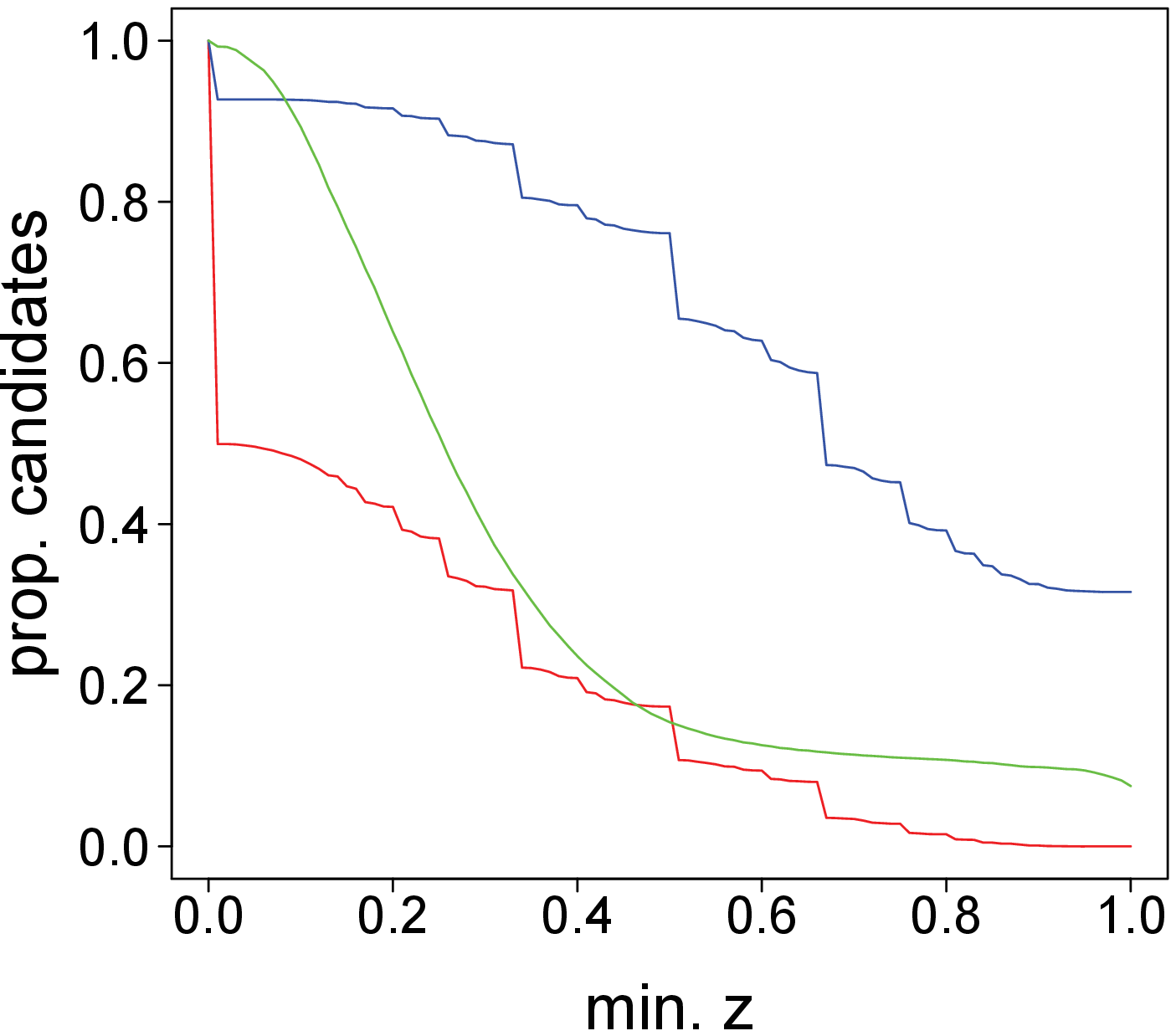}
\includegraphics[width=0.3\textwidth, trim=2cm 0.8cm -1cm 1cm, clip=true]{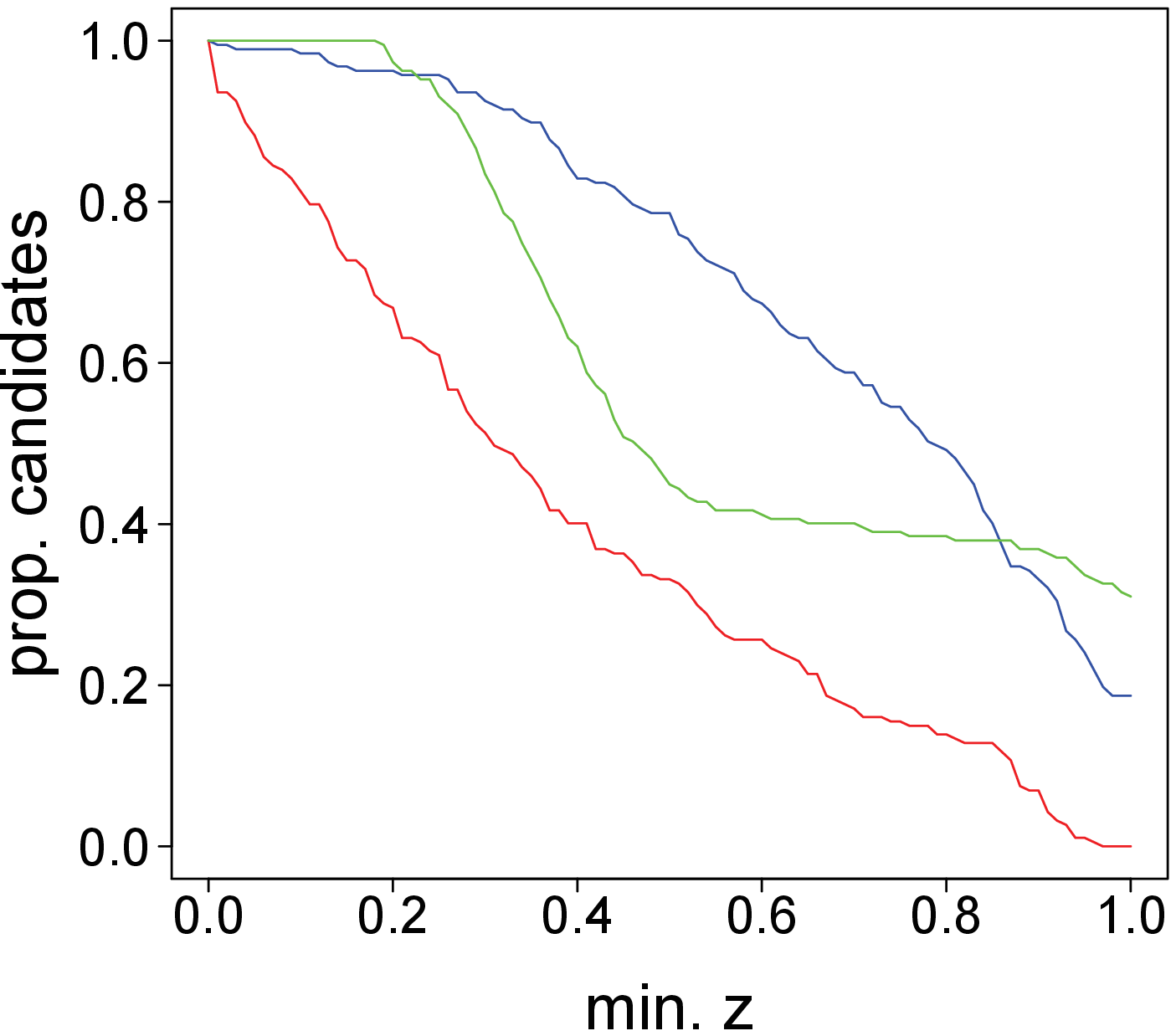}
\includegraphics[width=0.3\textwidth, trim=0cm 0.8cm 1cm 1cm, clip=true]{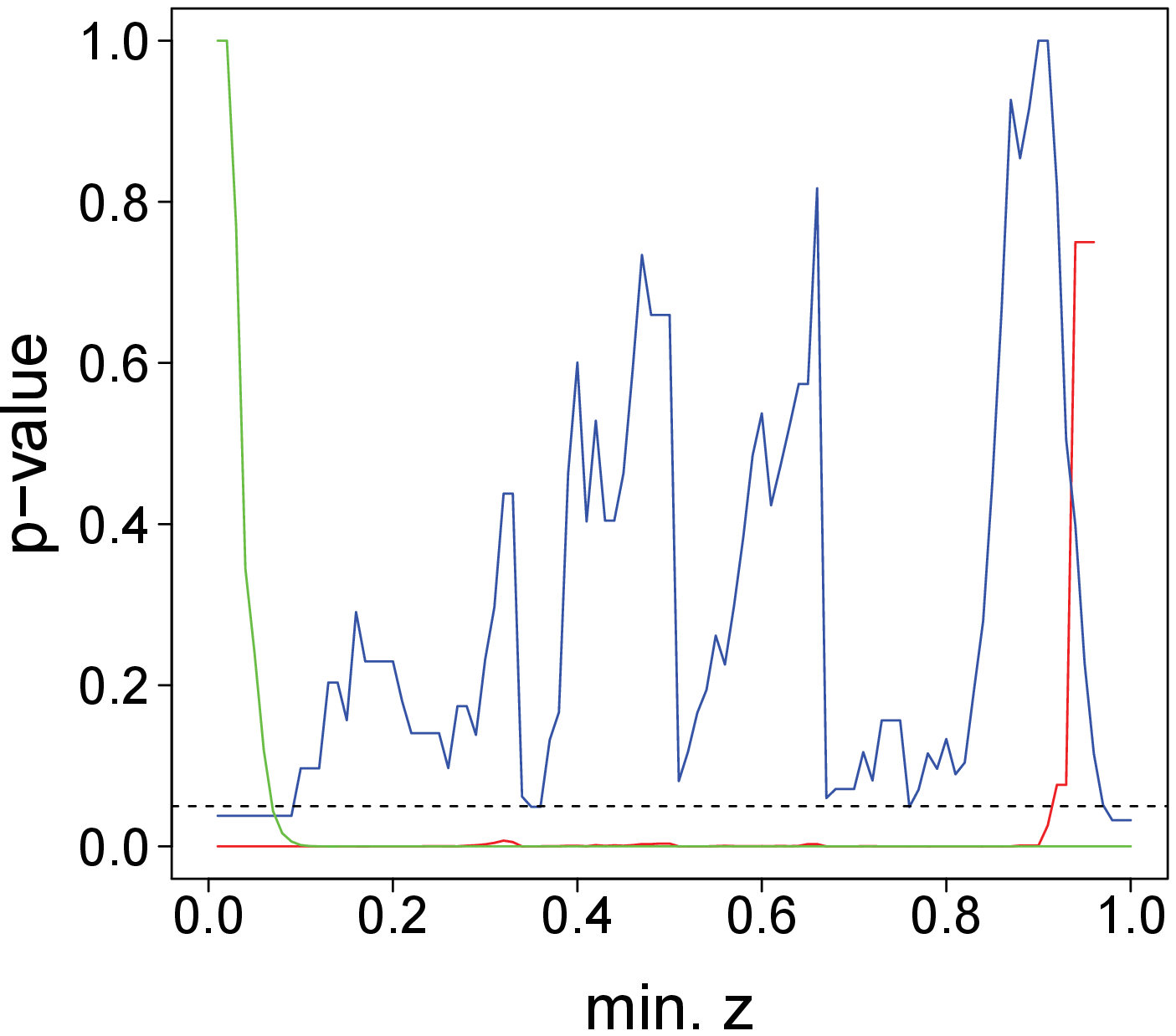}\\
\includegraphics[width=0.8\textwidth, trim=1cm 7.5cm 2.5cm 2cm, clip=true]{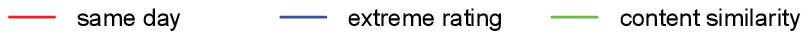}
\caption{Aggregate behaviour of random spammers (A) and candidate spammers identified using deviation from majority opinion (B), with p-values indicating the significance of observed differences between the two groups (C). Dashed line indicates a significance of 0.05. Features considered measure reviewers tendency to post reviews on the same day, use of extreme ratings, and degree of content similarity.}
\label{AmazonEvaluation}
\end{figure}

Figure \ref{AmazonEvaluation} shows the aggregate behaviour of reviewers for our set of alternative features, taken across a random sample of reviewers (panel A) and the 187 candidate spammers identified by our approach (panel B). Also shown is the p-values resulting from a proportional test, described in detail below. This p-value indicates the statistical significance of the differing behaviour between the random sample of reviewers and those reviewers identified as spammers using our approach.

To generate the random sample of reviewers, we uniformly sampled 100 random reviewers and then calculated scores and proportions within this sample. We reasoned that a sample size of 100 reviewers was a good representation of a set of reviewers identified by some process as high-quality candidate spammers. To get a reasonable estimate of the average behaviour, we repeated the sampling 100 times, and then took the mean proportions across each of these samples.

For each candidate we calculated three scores, giving the maximum content similarity between any two of their reviews, the proportion of reviews occurring with the same date and the proportion of reviews having extreme ratings. For each of the two groups, we then calculated the proportion of reviewers having scores greater then a threshold value $z$, for each of the respective scores. A proportional test was then applied for each value of $z$ to determine whether or not the observed differences between the two groups was statistically significant.

As shown in Figure \ref{AmazonEvaluation}, the proportion of reviewers that exhibit high content similarity is significantly increased amongst those reviewers identified using our approach. The proportion of reviewers publishing reviews on the same day is also significantly increased amongst these reviewers. Both groups exhibit similar behaviour with respect to use of extreme ratings, with significant differences only occurring for minimal or maximal values of $z$.\\

Following  from our analysis of aggregated behaviour, we also performed a more detailed investigation of the 20 candidates deemed to be the most likely spammers. Candidate spammers were ranked according to the proportion of non-majority reviews (reviewers with a greater proportion ranked higher) and the top 20 were selected for further investigation. Similar to our analysis of aggregate behaviour, this investigation considered review text, rating scores, product categories and the timing of posted reviews. Previous studies have shown that repeated text, numerous extreme ratings (1-star or 5-star), and posting of multiple reviews on the same day, particularly within a narrow range of highly similar products, are strong indicators of spam \citep{Jindal:2008,Mukherjee:2012,Mukherjee:2013,Akoglu:2013}. Therefore we took occurrences of these markers as evidence of opinion spam.

Table \ref{TopTwenty} describes the results of our investigation for the top 20 candidate spammers. We found that of these top 20, at least 17 were highly likely to be opinion spammers, having some combination of repeated text, numerous reviews posted on the same day, or large numbers of extreme reviews focused on particular product groups.\\

The average running time for our approach was 191 seconds (~5 million edges, average over 5 runs, single execution thread on an i5 2.4GHz processor with 8GB RAM). We note for comparison that FraudEagle is reported to require ~100 seconds to perform a single iteration over a graph containing approximately 1 million edges \citep{Akoglu:2013}. 

\begin{table*}[!ht]
\caption{Results of manual investigation of top 20 ranked reviewers. Reviewers were evaluated based on the number of extreme ratings (1-star or 5-star ratings) awarded, targeting of highly similary product groups (e.g. numerous 1$\star$ ratings for a particular author), presence of repeated text in multiple reviews (at least an entire sentence), and posting of multiple reviews within a short period of time. Reviewers marked with a star showed little or no corroborating evidence of spam.}
\begin{tabular}{l|l|l|l|l|l|l|p{4cm}}
Rank & \# reviews & \# 1$\star$ & \# 5$\star$ & Similarity & Rep. text & Rep. dates & Comment\\
\hline
1 & 42 & 40 & 2 & F & F & T & Sentiment of text doesn't match rating\\
2 &  37 & 35 & 2 & T & F & T &\\
3 &  27 & 27 & 0 & T & F & T &\\
4 &  42 & 40 & 0 & T & F & T & Clearly dislikes style of music, but numerous reviews\\
5 &  27 & 25 & 2 & T & F & T & Multiple (10) reviews of same product with different text\\
6 &  43 & 42 & 1 & T & T & T & \\
7 &  32 & 23 & 2 & F & F & T & \\
8 &  21 & 21 & 0 & T & F & T &\\
9 &  44 & 31 & 6 & F & F & T & Username is 'United Federation of Trolls', reviews no longer on Amazon website\\
10 &  34 & 29 & 3 & T & T & T &\\
11$^{*}$ &  31 & 16 & 3 & F & F & F &\\
12 &  42 & 16 & 0 & T & F & F & 1$\star$ only on recent pop albums, reviews no longer on Amazon website\\
13 &  39 & 29 & 6 & T & F & F & Verified purchase only on 5$\star$ reviews\\
14$^{*}$ &  53 & 11 & 2 & F & F & T &\\
15 &  31 & 22 & 9 & T & F & T\\
16 &  29 & 22 & 6 & T & T & T & Numerous 1$\star$ reviews directed at single musician\\
17 &  37 & 15 & 5 & T & T & T & Multiple reviews of same product\\
18$^{*}$ &  52 & 22 & 12 & F & T & F &\\
19 &  51 & 45 & 5 & T & T & T & Strong bias in review text\\
20 &  31 & 22 & 7 & T & F & T & Strong bias in review text\\
\hline
\end{tabular}
\label{TopTwenty}
\end{table*}

\section{Discussion and Conclusions}

Application of our approach to the Amazon data set shows the proportion of reviews disagreeing with the mean is a good indicator of spammer behaviour. Our investigation of those reviewers identified as candidate spammers showed strong evidence of spam behaviour. By considering specific features of the problem domain, namely the desire of spammers to drive mean ratings in a particular direction, we are able to identify how spammer behaviour differs from the behaviour of honest reviewers. Having identified this difference we are able to formulate a simple test to detect the relevant behaviour. As a result, a major advantage of our approach is its simplicity and the consequent minimal computational requirements. Our approach could easily be combined with other descriptors of spammer behaviour (e.g. timing of reviews \citep{Xie:2012,Fei:2013}, or, if available, text-based features) as part of a more complex classifier, such as those proposed in \citet{Mukherjee:2012,Mukherjee:2013}.

In undertaking our investigation of candidate spammers we noted an interesting phenomenon that we have not seen discussed in previous studies. Many of the reviews published by the top 20 spammers were highly derogatory and were clearly aimed at a particular subset within a given product range. For example, one reviewer had singled out 3 specific female pop artists, all having a similar style of music, and had written highly negative reviews for multiple albums by each of these artists. It is clear from the review text that this reviewer has an aversion to this particular style of music, and it seems highly unlikely that this reviewer had genuinely purchased each of the target albums expecting to enjoy them. Another reviewer displayed a strong political bias in their rating of books, while another claimed that a particular video game was the greatest game ever made and posted numerous 1$\star$ reviews of other games in the same genre. We find this type of reviewer to be quite interesting, as the review text in no way attempts to disguise the reviewers bias and pose as an objective review. We term this type of behaviour \textit{agenda spam}, whereby a particular reviewer appears to have a specific political or personal agenda and uses product reviews as a kind of public (yet anonymous) forum in which to express their views. Note that while these types of review may not be intended to derive financial benefit, we argue that they still represent a form of spam, as they pollute the review space with non-objective, highly-biased and, we suggest largely unhelpful, reviews.\\

While the approach presented in this paper is able to identify opinion spammers, there are a number of improvements we would like to consider in future work. First, our approach considers a global probability that a random review will differ from the mean rating. However, calculation of this global probability currently includes  spammer reviews. This probability could instead be treated as a hyper-parameter, being updated after running the iterative correction of the mean ratings. Second, we use the mean rating as a measure of majority opinion. In future work, we would like to consider alternative models of majority opinion, and more detailed descriptions of the distribution of rating scores over each product. Third, we believe that our approach should be combined with alternative approaches to gain a more comprehensive detection system. Spammers employ a wide range of strategies, and particular strategies may be more easily detected using a particular approach. Combining these different approaches in a flexible manner would provide an effective method for dealing with the multi-faceted nature of spammer behaviour.

In addition to the improvements outlined above, future work will also consider how the type of approach presented in this paper can be extended to detect groups of spammers acting in a coordinated manner. Previous work has shown that detection of spammer groups is possible by considering groups of reviewers having multiple products in common. We suggest that coordinated behaviour amongst reviewers should be relatively unusual, so that spammer groups may be identified based on their deviation from normal, uncoordinated behaviour.\\

Opinion spam is a continuing problem for consumers looking to be guided by online product reviews in making their purchasing decisions. In this paper we have demonstrated a simple method for detecting opinion spam based only on reviewer ratings. By applying a binomial test our characterisation of spammer behaviour can be used to detect opinion spammers in real data, with results showing improvements over existing methods. We suggest that the simplicity of our approach is extremely desirable given the volume of data typically considered for opinion spam and suggest that our characterisation could be easily combined with other text and non-text based features as part of a multi-criteria system.

\section*{Acknowledgements}

This research is supported in part by an Australian Research Council Linkage Project (LP120200128) and in part by the Australian Transaction Reports and Analysis Centre. We would like to thank Mr. Claude Colasante for his generous support for this research project. We would also like to thank the anonymous reviewers of this paper for their helpful comments.

\section*{References}
\bibliography{refs}

\end{document}